\documentclass[useAMS,usenatbib]{mn2e}
\usepackage{graphicx}
\usepackage{epsfig} 
\usepackage{multirow} 
\usepackage{aas_macros}

\title[FIR star-formation in XMMU J2235.3-2557]
{Dust-obscured  star-formation in the outskirts of \\
XMMU J2235.3-2557, a massive galaxy cluster at $z$=1.4\thanks{Based on observations with Herschel, an ESA space observatory with science 
instruments provided by European-led Principal Investigator consortia and with important participation from NASA.}  \\ }

\author[J.S. Santos et al.]
{J. S. Santos$^{1}$\thanks{E-mail:jsantos@sciops.esa.int}, B. Altieri$^{1}$, P. Popesso$^{2}$, V. Strazzullo$^{3}$, I. Valtchanov$^{1}$, S. Berta$^{2}$, \and
 H. B\"ohringer$^{2}$, L. Conversi$^{1}$, R. Demarco$^{4}$, A. C. Edge$^{5}$, C. Lidman$^{7}$, D. Lutz$^{2}$, \and 
L. Metcalfe$^{1}$, C.R. Mullis$^{7}$,  I. Pintos-Castro$^{8,9,10}$, M. S\'anchez-Portal$^{1}$, T. D. Rawle$^{1}$, \and 
P. Rosati$^{11}$, A. M. Swinbank$^{5}$ and M. Tanaka$^{12}$ \\
$^{1}$European Space Astronomy Centre (ESAC)/ESA, Villanueva de la Ca\~nada, 28691, Madrid, Spain\\
$^{2}$Max-Planck-Institut f\"ur extraterrestrische Physik, Giessenbachstra\ss e, 85748 Garching, Germany\\
$^{3}$CEA \/ Saclay, Service d'Astrophysique, L'Orme des Merisiers, B\^at. 709, 91191 Gif-sur-Yvette Cedex, France \\
$^{4}$Department of Astronomy, Universidad de Concepcion, Casila 160-C, Concepci—n, Chile\\
$^{5}$ Department of Physics, Durham University, Durham DH1 3LE \\
$^{6}$Australian Astronomical Observatory, PO Box 915, North Ryde NSW 1670, Australia \\
$^{7}$Wachovia Corporation, NC6740, 100 N. Main Street, Winston-Salem, NC27101\\
$^{8}$Instituto de Astrof\'isica de Canarias, La Laguna, Tenerife, Spain \\
$^{9}$Departamento de Astrof\'isica, Facultad de F\'isica, Universidad de La Laguna, La Laguna, Tenerife, Spain \\
$^{10}$Centro de Astrobiolog\'ia, INTA-CSIC, Villanueva de la Ca\~nada, Madrid, Spain \\
$^{11}$European Southern Observatory (ESO), Garching, Germany\\
$^{12}$Institute for the Physics and Mathematics of the Universe, The University of Tokyo, 5-1-5 Kashiwanoha, Kashiwa-shi, Chiba 277-8583, Japan\\  }

\begin{document}

\date{Accepted . Received; in original form}

\pagerange{\pageref{firstpage}--\pageref{lastpage}} \pubyear{2002}

\maketitle

\label{firstpage}

\begin{abstract}
Star-formation in the galaxy populations of local massive clusters is
reduced with respect to field galaxies, and tends to be suppressed
in the core region. Indications of a reversal of the
star-formation--density relation have been observed in a few $z >$1.4
clusters. Using deep imaging from 100-500$\mu$m from PACS and SPIRE
onboard Herschel, we investigate infrared properties of
spectroscopic and photo-$z$ cluster members, and of H$\alpha$ emitters in XMMU
J2235.3-2557, one of the most massive, distant, X-ray selected
clusters known. Our analysis is based mostly on fitting of the
galaxies spectral energy distribution in the rest-frame 8-1000$\mu$m. We
measure total IR luminosity, deriving star formation rates (SFRs)
ranging from 89--463 M$_\odot$/yr for 13 galaxies individually detected by
Herschel, all located beyond the core region ($r >$250 kpc). We
perform a stacking analysis of nine star-forming members not
detected by PACS, yielding a detection with SFR=48$\pm$16 M$_\odot$/yr. Using a
color criterion based on a star-forming\ galaxy SED at the
cluster redshift we select 41 PACS sources as candidate star-forming
cluster members. We characterize a population of highly obscured SF
galaxies in the outskirts of XMMU J2235.3-2557. We do not find
evidence for a reversal of the SF-density relation in this massive,
distant cluster.

\end{abstract}


\section{Introduction}

The high-density environments of galaxy clusters are ideal laboratories to study the formation and evolution of galaxies. 
A wide range of physical mechanisms takes place in these nodes of the cosmic web, namely galaxy harassment, tidal and ram pressure stripping and 
mergers, that result in a population of galaxies with a diverse range of properties. 
However, two distinct families stand out, the early-types or passive galaxies, and the late-types or star-forming galaxies.
While the cores of nearby massive galaxy clusters are typically dominated by red, passive galaxies with old stellar populations, 
star-forming galaxies are usually located at the cluster outskirts, where the quenching mechanisms are no longer dominant 
(e.g. Treu et al. 2003, Haines et al. 2007).

The high-mass end of the galaxy populations in clusters appears to be dominated by quiescent, early-type galaxies, 
even up to $z$ =1.4 \citep[e.g.][]{Lidman}. However, as we approach the global cosmic star-formation density peak (1$<z<$2), we expect an 
enhancement of star formation (SF) to take place in the biased environments of clusters, the so-called reversal of the SF--density relation \citep{Dressler}.
The identification of the epoch when these massive galaxies showed the first signs of SF
as a function of galaxy mass, environmental density and cluster halo mass can set tight
constraints on structure formation models \citep{Hopkins, Delucia}.
Still, the details (i.e., fraction and location of star forming galaxies and amount of SF) of this increase remain unclear. 

Most of the energy from star-formation and active galactic nuclei (AGN) activity is absorbed by dust and re-radiated in the infrared.
Prior to the advent of orbital infrared observatories such  as ISO, Akari and \textit{Spitzer}, one of the most widely used methods 
for determining SFRs in galaxies had been through measurements of the H$\alpha$  6563~\AA~luminosity \citep{Kennicutt, Kewley}.
However, extending this method to high-redshifts is challenging due to the difficulties of obtaining ground-based NIR spectroscopy of faint galaxies.
In the absence of H$\alpha$, star formation is usually quantified via the [OII] $\lambda$ 3727 emission line 
\citep{Heckman, Hayashi}, although a degree of uncertainty exists about the reliability of [OII] as a star 
formation tracer \citep{Coia,Lemaux}, as it is sensitive to dust extinction and metalicity, as well as being enhanced in LINERs.

Observations with ISO \citep{Coia, Metcalfe} and later on with MIPS onboard \textit{Spitzer} provided a significant development in understanding 
star-formation in galaxies at intermediate redshift.
\cite{Saintonge} observed a significant increase in the fraction of dusty star-forming cluster galaxies up to $z$ 
= 0.83 (now dubbed the MIR Butcher-Oemler effect). Mid-infrared observations of galaxy clusters at the highest known redshifts
($z\sim$1.5) have begun to reveal a population of IR luminous, actively star forming galaxies deep within the cluster cores \citep{Hilton, Tran}. 
In particular, \textit{Spitzer}/MIPS 24$\mu$m observations of the moderately massive cluster XCS J2215-1738 at $z$ = 1.46, 
show that this cluster hosts a significant population of galaxies with very high SFRs, even in core region \citep{Hilton}.  

The \textit{Herschel} observatory \citep{Pilbratt} is the largest space telescope to date and 
provides unrivalled sensitivity in the wavelength window ranging from 55 to 672$\mu$m. \textit{Herschel} brackets the 
critical peak of FIR emission of $z\sim$1-2 galaxies, providing a direct, unbiased measurement of star-formation. 
Several Herschel Key Projects were designed to measure the star-formation activity at 
high-redshift and to investigate the role of environment in regulating the star formation activity.
In particular, results from the Photodetector Array Camera and Spectrometer (PACS) Evolutionary Probe Key Project \citep [PEP,] []{Lutz} 
indicate that for massive galaxies ($M/M_\odot >$ 10$^{11}$) 
at $z\sim$1 the mean specific star formation rate tends to be higher at higher density \citep{Popesso11}.

Using recently acquired \textit{Herschel} imaging data, we present in this paper a detailed account of the dust-obscured 
star-formation properties in the galaxy population of the X-ray selected cluster XMMU J2235.3-2557 
(hereafter XMM2235) at $z$=1.393 \citep{Mullis}. This massive, distant cluster was 
discovered by the XMM-Newton Distant Cluster Project (see \cite{Fassbender} for a recent review of the survey results). 
Given its remarkable position in redshift and its high mass \citep{Jee}, XMM2235 has been the 
subject of deep, multi-wavelength follow-up observations using the major astronomical observatories. 
Near-infrared (NIR) J/Ks data obtained with HAWK-I revealed a tight red-sequence with a low-scatter 
in the J-Ks color-magnitude relation \citep{Lidman} and a clear segregation between the core, red galaxies, and the outer, bluish galaxies.
Using HST/ACS and NICMOS optical/near-IR data, the properties of the passive cluster galaxies have been accurately studied
 \citep{Strazzullo}, indicating that star formation appears to be effectively suppressed in the cluster center ($r<$ 250 kpc).
 Furthermore, the SED-derived star formation ages confirmed an old population of galaxies with $z_{form}>$3 in the inner core and a 
 significant younger population in the outskirts \citep{Rosati}. 
 These studies show that despite the lookback time of 9 Gyr, XMM2235 has an evolved galaxy population, with a massive  
brightest cluster galaxy (BCG). More recently, a narrow-band imaging survey searching for H$\alpha$ emitters \citep{Bauer, Grutzbauch} 
added further investigation on the star-formation properties of the cluster galaxy population. 

\begin{figure*}
\includegraphics[height=8.3cm,angle=0]{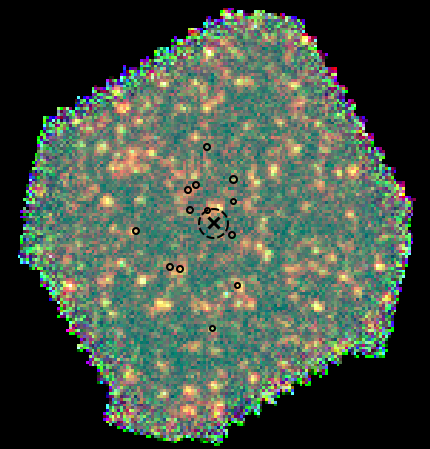} \hspace{2cm}
\includegraphics[height=8.3cm,angle=0]{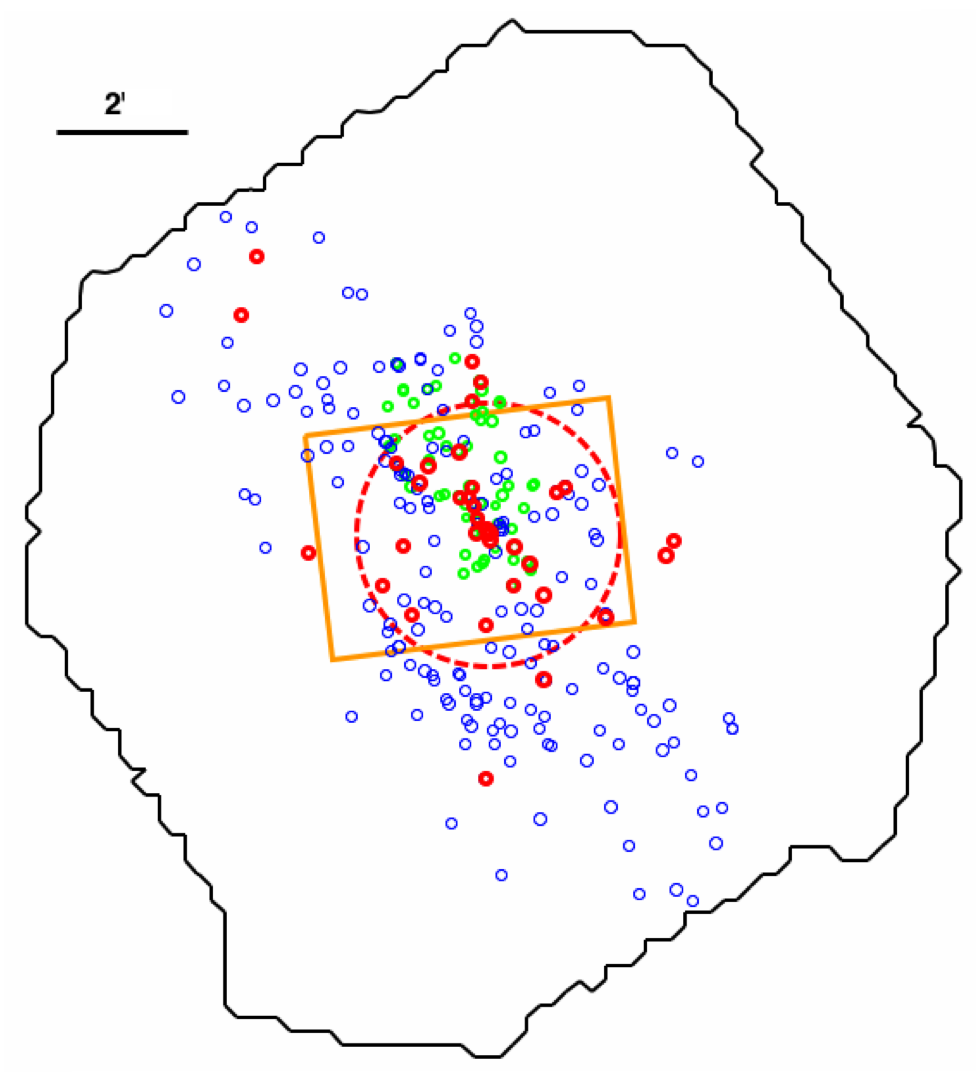}  
 \caption{\textit{Left}  PACS+SPIRE RGB (250/160/100$\mu$m) image of XMM2235 centered on the cluster X-ray center (black cross).
 The black circle indicates a radius of 250 kpc centered on the BCG. The 13 cluster members with FIR emission (see Table 2) are shown with small circles. 
 \textit{Right} PACS 100$\mu$m footprint (black contour) displaying the complete  
 spectroscopic catalog (blue), the 35 confirmed cluster members (red) and the 59 H$\alpha$ emitters (green). The area with photometric 
 redshifts in enclosed by the orange rectangle and the red dashed circle indicates a radius of 1 Mpc centered on the BCG. 
 The images have a size of 15.5$\arcmin \times$15.5 $\arcmin$. North is up and East is to the left. }
\end{figure*}

The paper is organized as follows: 
in \S 2 we describe the Photodetector Array Camera and Spectrometer (PACS) and Spectral and Photometric Imaging REceiver (SPIRE) 
observations and reduction procedures.
In \S 3 we describe the extensive multi-wavelength ancillary data used in this study.
We perform the SED fitting of individual detections in \S 4 to derive the far-infrared properties of the cluster 
galaxies, namely, the total FIR luminosity, $L_{IR}$, star-formation rate (SFR) and specific star formation rate (sSFR). 
In \S 5 we perform a stacking analysis of the non-detections and in \S 6 we present the sSFR--stellar mass relation.
The presence of X-ray AGNs in our sample and their luminosity is quantified in \S 7.
Using the high-resolution HST/ACS data we determine the morphologies of the cluster members with \textit{Herschel} detection in \S 8.
Based on the FIR SED shape of the sources we define a sample of PACS cluster candidates in \S 9.
In \S 10 we present the analysis of the integral field unit (IFU) SINFONI spectroscopy data of the BCG. 
Our conclusions are summarized in \S 11.  

The cosmological parameters used throughout the paper are: $H_{0}$=70 km/s/Mpc,
$\Omega_{\Lambda}$=0.7 and $\Omega_{\rm m}$=0.3. Magnitudes are reported in the AB system and we will use the 
Salpeter initial mass function (IMF) unless otherwise stated.

\begin{table}
\caption{Summary of the datasets used in our analysis. The "Type" column refers to spectroscopy (S) or imaging (I).}  
\label{table:3}     
\small
\centering           
\begin{tabular}{llll} 
\hline\hline   
Instrument          & Type    &        Observed Band                         &        Field-of-view                      \\
\hline                         
FORS2               &  S    &    R, z (MOS/MXU)              &    6.5$\arcmin \times$13$\arcmin$              \\
ACS                   &   I    &  F850LP                              &    3.5$\arcmin \times$3.5$\arcmin$              \\
NIRI                   &   I    &  H narrow                             &    3$\arcmin \times$4$\arcmin$                   \\
HAWK-I               &    I    &  J, Ks                                    &   13.5$\arcmin \times$13.5$\arcmin$            \\
SINFONI            &   S   &  H                                       &    8$\arcsec \times$8$\arcsec$             \\
IRAC                  &   I    &  3.6/4.5/5.8/8.0  $\mu$m      &    14.5$\arcmin \times$8$\arcmin$                \\
PACS                 &   I    &  100/160 $\mu$m                &    11$\arcmin \times$11$\arcmin$                  \\
SPIRE                &   I    &  250/350/500 $\mu$m          &   16$\arcmin \times$16$\arcmin$                    \\          
ACIS-S              &    I    &  0.5-10 keV                         &  8.5$\arcmin \times$9$\arcmin$           \\
\hline
\end{tabular}
\end{table}

\section[]{Herschel observations and data reduction}

\begin{figure*}
\includegraphics[height=15.cm,angle=0]{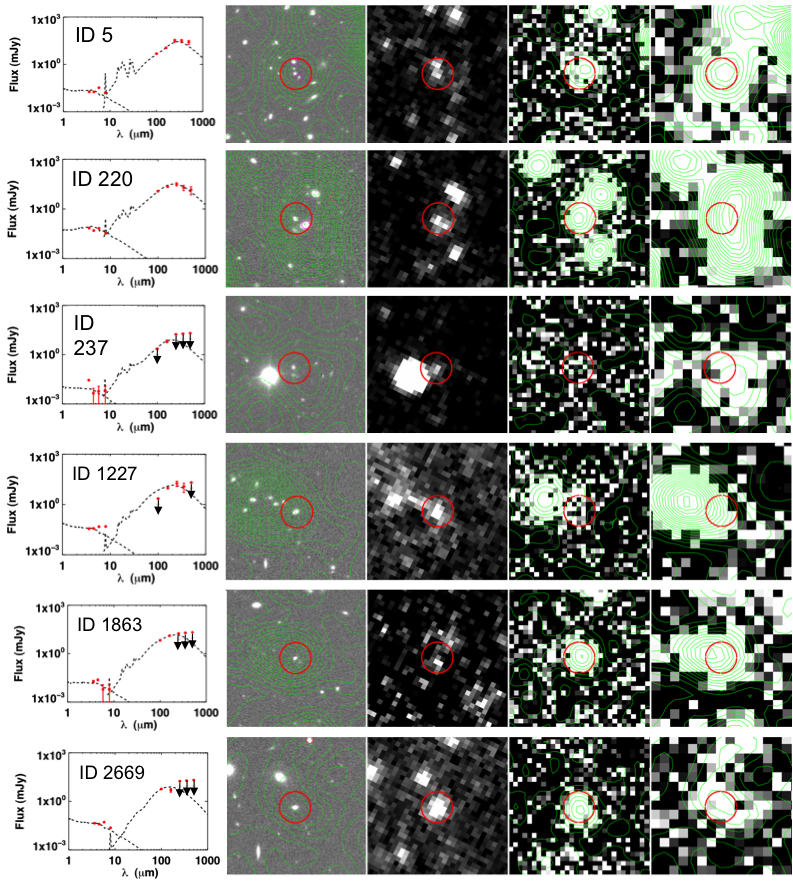} 
 \caption{Gallery of the 6 spectroscopically confirmed 
 cluster members with \textit{Herschel} detection. For each galaxy we show the FIR SED best-fit in the 
 observed frame, and stamps with 35$\arcsec \times$35$\arcsec$ of the K-band with 160$\mu$m contours overlaid in green, 8$\mu$m, 100$\mu$m and
 160$\mu$m band (ordered left to right). The red circles have a radius of 4$\arcsec$.  The exact positions of the PACS blind detection catalog used in the flux 
 measurements are not shown to avoid crowding.  }
\end{figure*}

\begin{figure*}
\includegraphics[height=15.cm,angle=0]{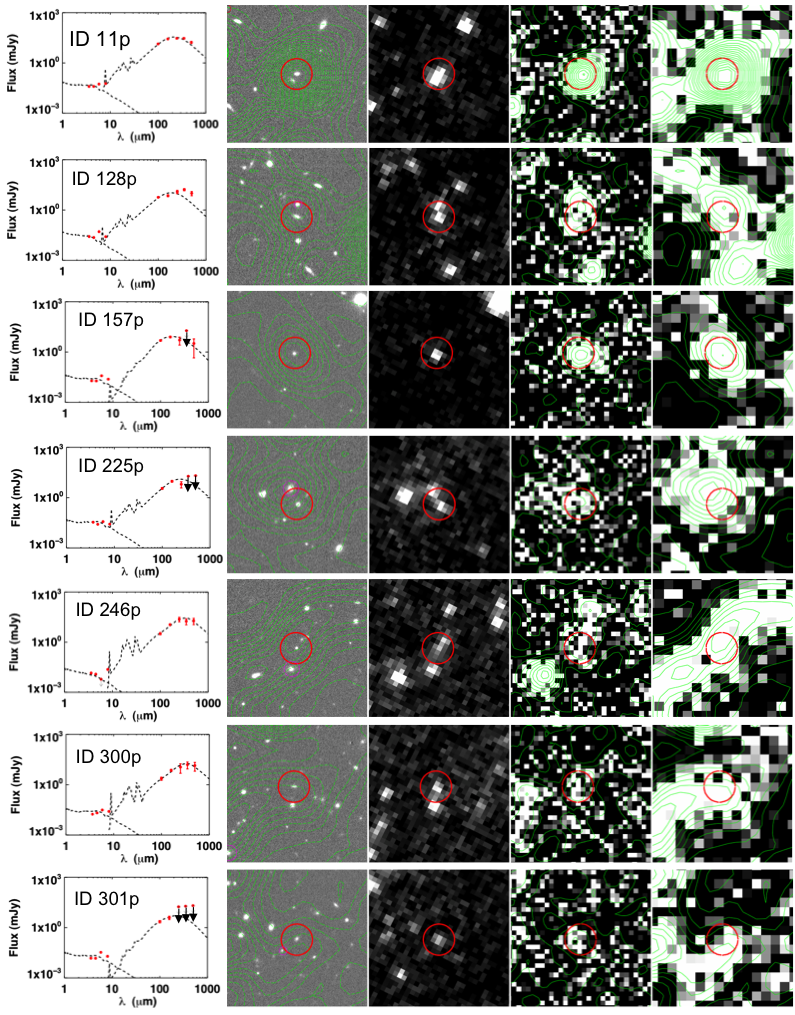} 
 \caption{Gallery of the 7 photometric redshift cluster members with \textit{Herschel} detection. For each galaxy we show the FIR SED best-fit in the 
 observed frame, and stamps with 35$\arcsec \times$35$\arcsec$ of the K-band with 160$\mu$m contours overlaid in green, 8$\mu$m, 100$\mu$m and
 160$\mu$m band (ordered left to right). The red circles have a radius of 4$\arcsec$.}
\end{figure*}

The \textit{Herschel} observations of XMM2235 were carried out as part of two programs, one in guaranteed time (GT1, PI Altieri) 
and one in open time 1 (OT1, PI Popesso).

The PACS 100/160$\mu$m imaging data were taken within the scope of the 100h OT1 program 
designed to study the FIR star formation history of a sample of 8 X-ray selected galaxy clusters at 
1.4 $< z <$ 1.8, and investigate the relation between star formation activity and environment at the epoch 
when clusters are assembling galaxies at a faster rate than today and galaxies are still undergoing their own formation process.
The results obtained for the complete cluster sample observed in this program will be presented in 
forthcoming papers. 

The SPIRE 250/350/500$\mu$m imaging data were taken in the context of the smaller and shallower GT1 
program, aimed at studying the evolution of star-formation in 8 high-redshift clusters within the 
range 0.8 $<z<$ 2.4, including two well-known proto-clusters \citep [see][] {Seymour}, Sanchez-Portal et al. in prep.).  
This program complements the one observed in OT1, by increasing sample statistics and extending it to higher redshift.

\subsection{PACS observations and catalog}

The PACS \citep{Poglitsch} observations were acquired in December 2011 (ObsId = 1342234105, 1342234106, 1342234430, 1342234431), amounting to 
a total of $\sim$24 h. 
The full data reduction of the data considered here in addition to the full dataset of the original PACS Program are described in Popesso et al. 
(2012 in prep.). Here we summarize the main steps. Data are processed through the standard PACS data reduction pipeline in the Hipe environment 
\citep{Ott}, with the addition of some custom procedures, aimed at removal of instrument artifacts.  These procedures comprise: iterative masking of 
the sources for avoiding residuals due to the high-pass filtering implemented in the PACS pipeline. Visual inspection of scan legs to flag and discard 
scan legs affected by interferences in the blue channel. Offsets and errors in the pointing of the \textit{Herschel} satellite are corrected by 
re-centering the data on a grid of known bright priors. 
The map coaddition and error estimation is performed through a weighted mean of several maps with the same WCS. The PACS catalogs at 100 and 
160 $\mu$m are created via a blind extraction using the Starfinder PSF-fitting code \citep{Diolaiti}. Flux reliability, incompleteness and spurious source 
fraction are estimated and tested via Monte Carlo simulations. Dedicated IDL procedures, described in Popesso et al. (2012) are used to check the 
effect of the high-pass filtering task on the source flux in different flux bins by adding fake sources on the PACS pixel timeline. See also \citet{Berta}, 
\citet{Lutz} for further details about source extraction and data reduction. The images have a pixel scale of 1.2$\arcsec$/pix and 2.4$\arcsec$/pix in 
the 100$\mu$m and 160$\mu$m bands, respectively. The PACS PSF at the nominal scan speed (20$\arcsec$/s) is 6.7$\arcsec$ at 
100$\mu$m and 11.3$\arcsec$ at 160 $\mu$m.

The PACS 3-$\sigma$ blind detection catalogs have 344 in the 100$\mu$m band and 298 sources in the 160$\mu$m band, with 
corresponding 3-$\sigma$ sensitivity limits of 2.25 mJy and 4.5 mJy, respectively.
We cross-matched the two blind catalogs using a nearest neighbor algorithm with a cross-match distance of 5$\arcsec$ to obtain 
a catalog with 158 sources that are present in both PACS bands.  We use this catalog as prior for extracting sources in the SPIRE images.

\subsection{SPIRE observations and catalog}

The SPIRE \citep{Griffin} observations (ObsID: 1342210530, 34 min exposure) were performed using a cross-scan Large Map 
mode over a $10\times10$ arcmin region with 4 repetitions, achieving 1-$\sigma$ instrumental noise (3.5, 4.8, 5.4) mJy at (250, 350, 500) 
$\mu$m --- slightly below the SPIRE extragalactic confusion noise, (5.8, 6.3, 6.8) mJy at (250, 350, 500) $\mu$m \citep{Nguyen} in the three SPIRE bands. 
The standard SPIRE map making was used to generate the maps from the SPIRE bolometers timelines. 
To get SPIRE photometry for sources in XMM2235 we used Sussextractor 10.0.836, a source detection method developed for 
SPIRE maps \citep{Smith} to produced a catalog based on priors from PACS. 
For non-detections, we consider 3-$\sigma$ upper limits of 17.4 mJy, 18.9 mJy and 20.4 mJy at 250, 350 and 500 $\mu$m, respectively, 
including confusion noise and the flux calibration uncertainties.
The beam FWHM on our standard maps are (18, 25, 36) arcsec at (250, 350, 500) $\mu$m.

Using the 3-$\sigma$ flux limits in all 5 \textit{Herschel} bands, we obtain an estimate of the lower limit on the star-formation rate for these 
observations at $z$=1.4. We thus obtain a 3-$\sigma$ SFR equal to 105 M$_\odot$/yr, based on the \textit{Herschel} observations for this cluster.

\section{Ancillary multi-wavelength data}

The galaxy cluster XMM2235 benefits from an extensive multi-wavelength coverage, including deep, high-resolution 
X-ray data from \textit{Chandra}, optical imaging with the Hubble Space Telescope, near-infrared (NIR) imaging with VLT/HAWK-I 
and spectroscopic data with VLT/FORS2 taken over a large field-of-view. Details on the optical and NIR data used here can be found in \citet{Strazzullo} 
and \citet{Lidman} respectively. The high-quality, high-resolution Ks-band data obtained over a large FOV (13.5$\arcmin \times$13.5$\arcmin$) 
nearly matches the area covered with PACS.
This broad dataset is fundamental to the analysis of the \textit{Herschel} data 
which suffer from blending, and confusion only in the longest wavelengths.

\begin{table*}
\caption{Properties of the spectroscopically confirmed cluster members (first six rows) and photometric redshift members (seven rows after line break 
with suffix 'p' in ID) with \textit{Herschel} detection. To measure the projected radial distance of the photometric members from the central galaxy in \textit{kpc}, 
we assigned to those galaxies the cluster redshift, $z$=1.393.  }  
\label{table:2}     
\small
\centering           
\begin{tabular}{lllllllll} 
\hline\hline          
 ID      &   RA                      & DEC           	& $z$            & Stellar Mass                     &   $L_{IR}$                            & SFR                          & Distance  & [OII] / H$\alpha$ \\   
           &   (J2000)               &   (J2000)    	 &            	& ($\times$10$^{10}$M$_\odot$)  &  ($\times 10^{12} L_\odot$)   &   (M$_\odot$/yr)       & (kpc)            &        \\
\hline                        
%1	    & 338.78652    & -25.966369      &  1.3916            &     3.0$\pm$0.2                &     0.57$\pm$0.02    &        99$\pm$2       &      1500        &  no / no         \\
5*	    & 338.8407    & -25.954053      &  1.3986            &     2.0$\pm$0.1                &     1.55$\pm$0.19    &       268$\pm$34    &        250        &   yes / yes     \\
%9          & 338.84103   & -25.927588      &  1.3909            &     7.2$\pm$0.2                &     1.18$\pm$0.08    &      203$\pm$15     &      1030         &   yes / yes      \\
220 	    & 338.8212      & -25.997889      &  1.3879            &      4.8$\pm$0.2               &     2.47$\pm$0.32    &      427$\pm$55     &     1210         &   yes / --      \\
%231	    &  338.80357   &  -25.982125     &  1.3883            &     1.5$\pm$0.2                &     2.21$\pm$0.12    &     382$\pm$28      &     1190         &   yes / no         \\
237*	    & 338.8253    &  -25.968533     &  1.3816            &     3.3$\pm$0.2                &     0.57$\pm$0.01    &       99$\pm$2        &       414         &   yes / yes       \\
1227*   & 338.8874    &  -25.965951     &  1.3896            &     5.6$\pm$0.2                &     1.27$\pm$0.14     &     219$\pm$25      &     1541         &   no / --          \\
1863   & 338.8376      &  -26.022967     &  1.3972            &     3.0$\pm$0.1                &     1.28$\pm$0.38     &     220$\pm$66      &    1880         &    --  / --          \\
2669   & 338.8413    &  -25.917121     &  1.3902            &     6.2$\pm$0.2                &     0.77$\pm$0.18     &     131$\pm$33      &     1340         &    no / **         \\
\hline  
11p	    & 338.8588      &  -25.988553      &   1.38$\pm$0.16      &    4.9$\pm$0.2      &      2.71$\pm$0.26      &   463$\pm$44        &  1062        &  --  / --     \\
128p*   & 338.8651    &  -25.987018      &   1.38$\pm$0.24      &    5.3$\pm$0.2      &      1.03$\pm$0.16      &  176$\pm$26         &   1165       &   -- / --      \\
157p   & 338.8236    &  -25.948917      &   1.58$\pm$0.19      &    2.5$\pm$0.2      &      1.05$\pm$0.16      &  180$\pm$29         &   545         &   -- / yes     \\
225p*   & 338.8233    &  -25.935927      &   1.64$\pm$0.13      &    2.5$\pm$0.2      &      1.35$\pm$0.31      &  231$\pm$54         &   869         &   -- / **       \\
246p   & 338.8534    &   -25.942159     &   1.30$\pm$0.06      &    1.8$\pm$0.1      &      1.16$\pm$0.22     &   199$\pm$38         &   754         &   -- /  yes    \\
300p   & 338.8485    &  -25.938936      &   1.54$\pm$0.15      &    2.1$\pm$0.2      &     1.16$\pm$0.33      &   198$\pm$56         &   752         &   -- / yes     \\
301p*   & 338.8519    &   -25.953788     &   1.50$\pm$0.15      &    2.0$\pm$0.2      &     0.52$\pm$0.21      &    89$\pm$35         &   508         &   -- / **        \\
\hline
\end{tabular}
\flushleft  \hspace{0.3cm}   * FIR emission likely contaminated by neighbor galaxies
\flushleft  \hspace{0.3cm}   ** Not in the H$\alpha$ catalog but the galaxy is at the edge of the NIRI FOV
\end{table*}

For our analysis we use an extensive spectroscopic campaign based on 4 VLT/FORS2 masks using both the MOS and MXU modes, described in \cite{Rosati}. 
The catalog with 209 spectroscopic redshifts will be published in Strazzullo et al. (in preparation). 
The targets were initially selected using colors and magnitudes aiming at the detection of red members \citep{Mullis}.
Fainter objects with colors consistent with late and early type galaxies at the cluster redshift were targeted
in subsequent masks \citep{Rosati}. 
To date, a total of 35 cluster members have been confirmed within a redshift range 1.374 $< z <$1.406, and about a third of these 
(11) have a clear detection of the [OII] $\lambda$ 3727 emission line.

In addition to the spectroscopic catalog we also use photometric redshifts computed with the U,R,i,z,J,H,K bands, to overcome any bias or 
incompleteness of the spectroscopic sample and thus consider all galaxies which might be cluster members.
The photo-$z$ catalog encompasses the central $\sim$5$\arcmin\times$3$\arcmin$, where we have optimal optical/near-IR photometric coverage. 
The scatter of these photometric redshifts is $\delta z$/(1+$z$) $<$6\%, with an estimated contamination by interlopers of 50\% (obtained by 
comparison with the spectroscopic catalog). The photometric redshift range used for selecting cluster members is 1.1 $\le z \le$ 1.7. 
All the details concerning this catalog are found in \citet{Strazzullo}.

We also use here the recently published catalog of 163 narrow-band emitters detected in 2 narrow-band imaging pointings 
with NIRI on GEMINI/North, covering mostly the central part (3.5$\arcmin \times$5$\arcmin$) of the cluster \citep{Grutzbauch}. 
 The H narrow-band filter centered at a wavelength of 1.57 $\mu$m and with a width of 232~\AA~allows for the detection of the H$\alpha$ $\lambda$ 
 6563 line in the redshift range [1.372 -- 1.408].
 To ensure a robust analysis of these sources we opted to use only those which have an H$\alpha$ derived SFR $\ge$ 5.5 M$\odot$/yr, the 5-$\sigma$ 
upper limit quoted by the authors computed with an H$\alpha$ extinction of 1 magnitude. 
A total of 59 H$\alpha$ sources fulfill this criterion (see footprint of the narrow-band sources in Fig. 1) and will be used 
in our analysis. Using our extensive spectroscopic and photo-$z$ catalogs we find that twenty-five narrow-band emitters are cluster members 
(of which 10 have spectroscopic redshift), and 14 are interlopers (of which 7 have spectroscopic redshift). For the remaining 
20 narrow-band emitters 
we do not have redshift information. We would expect a higher correspondence between the H$\alpha$ galaxy candidates and our FIR data, 
relative to the  spectroscopically confirmed members, since the former are mostly star-forming by definition. There is, nonetheless, the possibility 
that extinction plays a prominent role, preventing highly star-forming galaxies to be seen in the narrow-band observations.

We use the \textit{Spitzer}/IRAC data taken in 2005 (PI C.R. Mullis, prog ID 20760), which is important to ensure a proper identification 
of the sources seen in \textit{Herschel} and link them to the optical/near-IR. 
The four-channel data were retrieved from the archive and reduced with standard procedures. 
Source extraction was performed in Point Response Function (PRF) multiframe mode within APEX. The PRFs used in the process were 
the default ones provided by the 
 Spitzer Science Centre. The data processing was performed with the
tool {\tt mopex} 18.5 using the standard parameters included in the {\tt mopex} pipeline provided by the Spitzer helpdesk.
We perform blind source extraction for each band and PSF-fitting photometry using {\tt mopex}. Sources  
below the 5-$\sigma$ detection thresholds ([0.0044, 0.0045, 0.0060, 0.0062] mJy at [3.6, 4.5, 5.8, 8] $\mu$m, respectively)  were discarded.

XMM2235 was observed with the \textit{Chandra} ACIS--S detector in VFAINT mode with a total effective exposure time of 196 ks 
(PI Mullis). The full data analysis and reduction were published in \citet{Rosati}.
The subarcsecond resolution of \textit{Chandra} allows for a robust detection of point sources, and therefore it is important for our work in
identifying AGN contamination in members with far-infrared emission (see \S 7). 

Lastly, we describe and analyze in Section 8 the deep IFU spectroscopy data of the brightest cluster galaxy obtained with 
SINFONI at the VLT, aimed at detecting star-formation in the BCG through the detection of the H$\alpha$ emission line.
In Table 1 we summarize the data used in this paper.

\begin{table*}
\caption{\textit{Herschel} fluxes in mJy of the spectroscopically confirmed cluster members (first six rows) and photometric redshift members 
(seven rows after line break with suffix 'p' in ID) with \textit{Herschel} detection. The absence of error bars indicates the value is an upper limit.}  
\label{table:2}     
\small
\centering           
\begin{tabular}{lllllllll} 
\hline\hline          
 ID      &   F$_{100}$   &  F$_{160}$   & F$_{250}$   & F$_{350}$   & F$_{500}$      \\   
\hline                        
5	    &    4.80$\pm$0.34         &   11.14$\pm$1.19    &  31.94$\pm$6.84   &    30.79$\pm$7.18        &    25.71$\pm$7.61              \\
220 	    &    11.77$\pm$0.59       &   22.00$\pm$1.00   &    31.31$\pm$6.69  &     19.69$\pm$6.94       &    14.51$\pm$7.41              \\
237     &     2.25$\pm$2.25         &    6.62$\pm$1.31     &     17.4                       &	    18.9                          &     20.4               \\
1227   &       2.25                          &        9.6$\pm$1.06   &    18.27$\pm$6.44    &      11.94$\pm$6.74     &    20.4         \\
1863   &     6.6$\pm$0.56           &    12.72$\pm$0.9     &     17.4    	             &     18.9	                    &    20.4         \\
2669   &    5.82$\pm$0.45          &      4.9$\pm$1.24     &     17.4     	             &     18.9	                    &    20.4          \\
\hline  
11p	    &   13.6$\pm$0.47          &    26.91$\pm$1.24   &    29.40$\pm$6.58   &   29.35$\pm$7.08      &    16.90$\pm$7.34          \\
128p   &    5.72$\pm$0.25          &   7.1$\pm$0.93        &    12.16$\pm$6.45    &  16.425$\pm$6.85     &     9.88$\pm$7.38         \\
157p   &    4.92$\pm$0.47          &   7.95$\pm$0.9        &    17.4              &   18.9                            &    20.4      \\
225p   &    3.57$\pm$0.47          &   9.45$\pm$1.21      &    17.4              &   18.9                            &    20.4                  \\
246p   &    3.22$\pm$0.45          &  11.01$\pm$0.88    &     23.07$\pm$6.69     &  17.54$\pm$6.83      &    18.48$\pm$7.41        \\
300p   &   2.33$\pm$0.50            &   7.21$\pm$1.07     &    11.21$\pm$6.40      &  15.60$\pm$6.85      &    13.85$\pm$7.46         \\
301p   &   2.27$\pm$0.36            &   3.89$\pm$0.88     &    17.4                            &   18.9                          &     20.4                     \\
\hline
\end{tabular}
\end{table*}

\section{FIR SED fitting of individual detections}

In this section we perform a standard SED-fitting of the infrared range with the widely used SED fitting code 
\textit{LePhare} \citep{Arnouts, Ilbert} that is based on a $\chi^{2}$ minimization. 
We use the suite of infrared galaxy templates published by \citet{Chary} (CE01) for the rest-frame 8-1000$\mu$m range, 
and \citet{Bruzual} templates (BC03) for the near-infrared part of the spectra.

To ensure a proper identification of the cluster members in the \textit{Herschel} maps we first performed a visual identification of 
all member galaxies in the Ks, IRAC and PACS images (see Fig. 2, 3). 
We then matched the PACS catalog with the spectro-photometric catalogs using a nearest match algorithm with a search 
radius of 5$\arcsec$. Blending affects some of the detections making a 
precise decontamination too difficult. In such cases, we take the following considerations to evaluate whether the FIR emission 
originates from the cluster member: (i) check whether the cluster member has [OII] or H$\alpha$ emission as an indication of SF, 
(ii) investigate the properties of the contaminants using our spectro-photometric catalogs, (iii) evaluate the spatial separation between 
the target galaxy and contaminant and how the relative emission of the galaxies evolves from the K-band, 8$\mu$m, 100$\mu$m to 
the 160$\mu$m images (see Fig. 2, 3). 
Based on these considerations we rejected 3 of 9 spectroscopic members detected either in the matched 100-160$\mu$m catalog 
or in the single band detections for which we performed manual photometry. 
In addition to the spectroscopic members, we also investigate the FIR properties of the galaxies with photometric 
redshift consistent with the cluster redshift, i.e., within 2-$\sigma$ of the cluster redshift (1.1 $\le z \le$ 1.7). 
Following the same procedure, we find a total of eight photo-$z$ members in the 3-$\sigma$ PACS catalog (see Tables 2, 3). 
The galaxy ID 5 is common to both the spectroscopic and photometric redshift catalog. 
With the exception of this cluster member, all sources lie at a distance greater than 250 kpc (equal to a third of $R_{500}$) 
from the cluster X-ray center. Half of the spectroscopic members show the [OII] signature in their spectra and the four cluster members detected by PACS with ID 5, 237, 1227 and 2669 have X-ray emission indicating AGN activity (see Section 7).
As indicated in Table 2, the FIR emission assigned to some cluster galaxies may be contaminated by background or foreground galaxies. In such cases, the measured SFR should be interpreted as an upper limit, since only dedicated near-infrared spectroscopy will 
allow us to properly identify the the star-forming galaxies. 

 \begin{figure}
\includegraphics[height=6.cm,angle=0]{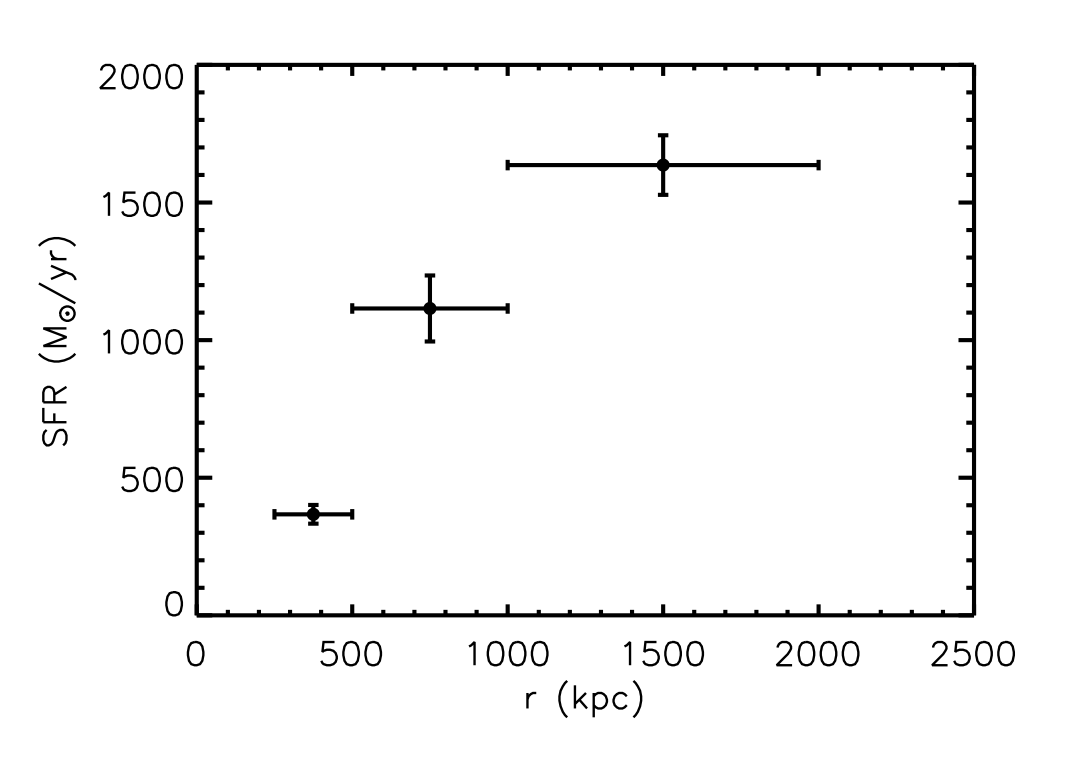} 
 \caption{SFR vs radial distance from the cluster center for the 13 cluster galaxies, distributed in 3 radial bins.}
\end{figure}

The total IR luminosities, $L(8-1000 \mu m)$ of the thirteen individual galaxy members are computed with \textit{LePhare} by integrating the best-fit SED models.
Using \citet{Kennicutt} that assumes a Salpeter IMF, we obtain the star-formation rate with the following conversion:
\begin{equation}
  SFR\,(M_\odot yr^{-1})=4.5\times10^{-44} L_{IR}\, (\textrm{erg s}^{-1})
\end{equation}

Luminous infrared galaxies (LIRGs) are defined as having 10$^{11}< L_{IR} / L_\odot <$10$^{12}$  and Ultra luminous infrared galaxies 
(ULIRGs) have a luminosity greater than 10$^{12}$ L$_\odot$. 
The cluster galaxies cover the range of infrared luminosities  0.5$\times$10$^{12}< L_{IR} / L_\odot <$ 2.7$\times$10$^{12}$, with about one third 
of the galaxies in the LIRG range and the remaining 2/3 are ULIRGS. 
In comparison with earlier work on 5 nearby, massive clusters with ISO data \citep{Metcalfe} in which no ULIRG was found among 35 cluster member 
detections, and more recent work (Haines et al. submitted) using Herschel observations of a larger sample of 30 clusters with only 1 to 2 detected ULIRGs, 
the fraction of ULIRGs in this distant cluster suggests a significant increase in the number of ULIRGs with redshift.

We measure star-formation rates in the range 89$\pm$35 $<$ SFR (M$_\odot$/yr) $<$ 463$\pm$44  for our 13 spectroscopic and photometric 
redshift cluster members. To investigate the spatial distribution of the SF members we sum the galaxies SFRs binned in 3 radial annuli: 
250 $< r_{1}<$ 500 kpc, 500 $< r_{2}<$1 Mpc, $r_{3}>$ 1 Mpc. As shown in Fig. 4, most of the cluster far-infrared star-formation occurs at a 
cluster centric radius larger than 1 Mpc, indicating that the cluster galaxy infaliing regions are the most active locations in this cluster.

\section{Stacking analysis}

In addition to the study of the cluster galaxies individually detected by PACS, we also constrained the star-formation of the cluster members that are 
potentially star-forming but do not have a detection in the FIR, as has been done in e.g., \citet{Dole}. 
We performed a mean stacking analysis using the IDL stacking software of \citet{Bethermin} and \citet{Bavouzet}  
on the residual maps at 100$\mu$m and 160$\mu$m  of two samples without PACS detection : 
(1)  the spectroscopically confirmed members with indications of on-going star-formation, 
and (2) the narrow-band emitters with a spectroscopic or photometric redshift within 2-$\sigma$ from the cluster redshift.
We tested the stacking procedure with simulations of dozens of gaussian sources with fluxes below the 3-$\sigma$ limit randomly placed 
in the residual maps and we recovered, within a very good agreement, the input fluxes, therefore we find the method to be robust.

\subsection{Non-detected, star-forming cluster members}

To select the cluster members that are potentially star-forming  
we require that the individually undetected galaxies fulfill one of the three conditions: (i) show the [OII] emission line, (ii) 
have a late-type morphology, and/or (iii) do not have red color. We end up with 9 potentially star-forming members not detected in 
either of the PACS bands.
  
 The stacked flux in the 100$\mu$m band is 0.65 mJy, a 2.6$\sigma$ detection considering the new 3-$\sigma$ detection limit of 3$\times$0.75/sqrt(N)= 0.75 mJy. 
 In the 160$\mu$m band, we obtain a 3$\sigma$ stacked flux of 1.6 mJy. The significance of these detections was tested with 1000 stacks of 9 randomly placed 
 sources in the residual maps. The average and standard deviation of the 1000 fluxes in both bands are a factor 2--3 lower than our stacked flux measurements, 
 therefore we are confident this is a real detection. 
Even tough there is an uncertainty as to which galaxies the signal may come from, this method allows us to go below the nominal star-formation threshold of the 
observations and place an upper limit to the average star-formation of this sample of individually undetected sources. 
As expected, the stacking signal is stronger in the 160$\mu$m (see Fig. 5) than at 100$\mu$m, as the former is more sensitive 
 to star-formation (i.e., closer to the SED peak) at the cluster redshift. We derive a star-formation rate for the stacked sources of 
 48$\pm$16 M$_\odot$/yr, which is about half of the 3$\sigma$ SFR limit for this cluster with the current data.

 \begin{figure}
 \hspace{1,0cm}\includegraphics[height=5.cm,angle=0]{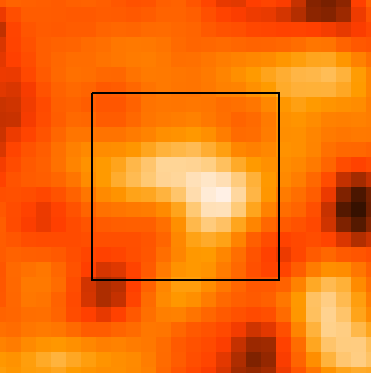} 
 \caption{PACS 160$\mu$m 1$\arcmin \times$1$\arcmin$, smoothed (gaussian kernel of 7\arcsec) image of the stacked 9 cluster members 
 that individually are not detected by \textit{Herschel}. The inner square has 30$\arcsec \times$30$\arcsec$.}
\end{figure}

\subsection{Non-detected narrow-band emitters}

We have 25 narrow-band emitters with either a spectroscopic or a photometric redshift in accordance with the cluster redshift.
In both the 100 and 160$\mu$m maps we have 20 non-detections of narrow-band emitters. 
At 100$\mu$m the stacked flux is much below the 3-$\sigma$ limit obtained for the stacked image, which is 0.50 mJy.
Similarly, in the 160$\mu$m image the corresponding stacked flux is a factor eight below the 3-$\sigma$ detection which is 1.01 mJy.
Based on these values we place an upper limit on the star-formation of the stacked non-detection of 24 M$_\odot$/yr,  which is 
consistent with the average SFR derived from the H$\alpha$ fluxes considering 2 magnitudes in extinction.

\section{The sSFR-$M_{*}$ relation}

The star-formation rate and specific SFR (sSFR) have been shown to correlate with the galaxies stellar mass ($M_{*}$).
These relationships provide clues to understand the origin of the reversal of the star formation-density relation, 
observed to occur at $z\sim$1 \citep{Elbaz07} in which distant galaxies show much larger typical SFR in
dense environments (i.e. group range) relative to low density environments. In this sub-section we investigate how the SFR, sSFR and $M_{*}$ correlate in XMM2235. 

To compute the galaxy stellar masses we wish to probe the rest-frame 1.6$\mu$m bump \citep{Gavazzi}. 
At the cluster redshift, the 3.6$\mu$m IRAC band corresponds to rest-frame 1.5$\mu$m and is therefore the closest to the 1.6$\mu$m 
emission. However, some of our galaxies are affected by blending in the IRAC
bands, therefore we derive the stellar masses by scaling the observed Ks-band magnitudes using the conversion  

\begin{equation}
M_{*}= (10.^{17.75Ê - Ks \times 0.349}) / 0.62 \,\,\,\,\,\,\,\,\, [M_\odot]
\end{equation}
\noindent that was determined in \citet{Strazzullo} to convert the K-band luminosity function of blue sources (based on an appropriate color selection) 
to the stellar mass function. We note that the calibration of the 3.6$\mu$m magnitude vs stellar mass follows this relation very closely.

The specific star-formation rate, which is simply the SFR divided by the stellar mass, measures the star-formation efficiency of a 
galaxy and the fraction of the galaxy mass that can be converted into starlight per unit time. It has been shown that the sSFR increases 
with redshift independently of mass and the sSFR of massive galaxies is lower at all redshifts \citep{Perez}. Still, a 
degree of uncertainty and controversy exists in the details (slope and scatter) of these relations.

\citet{Elbaz11} claim that  high-redshift ULIRGS form stars in the so-called "normal" main-sequence 
mode \citep[see also][] {Noeske}, in contrast to the local ULIRGs that are mostly starbursts. The proposed 
redshift evolution of the sSFR of main sequence galaxies is parametrized in the following way: 
\begin{equation}
sSFR_{MS}\,\, [Gyr^{-1}] = 26 \times t_{cosmic}^{-2.2}  
\end{equation}

For XMM2235, the age of the Universe at the cluster redshift is 4.62 Gyr which implies a sSFR$_{MS}$= 0.9 Gyr$^{-1}$. 
We draw this line in Figure 6 and find that all of our individual detections are indeed above the main-sequence line in the shaded region that 
corresponds to Eq. (14) of \citet{Elbaz11} which defines the sSFR of a starburst as $>$ 52 $\times t_{cosmic}^{-2.2}$. 
We ascribe this effect purely to a selection bias, as the SFR limit of these observations corresponds 
to 70(2-$\sigma$)--105(3-$\sigma$) M$_\odot$/yr, as indicated by the red dashed line in Fig. 6. This means that with the depth of these \textit{Herschel} 
observations we can only detect sources above the main-sequence (MS). Only by stacking sources we can reach the MS level (see the magenta star 
symbol in Fig. 6).

\begin{figure}
\hspace{-0.5cm}\includegraphics[height=6.5cm,angle=0]{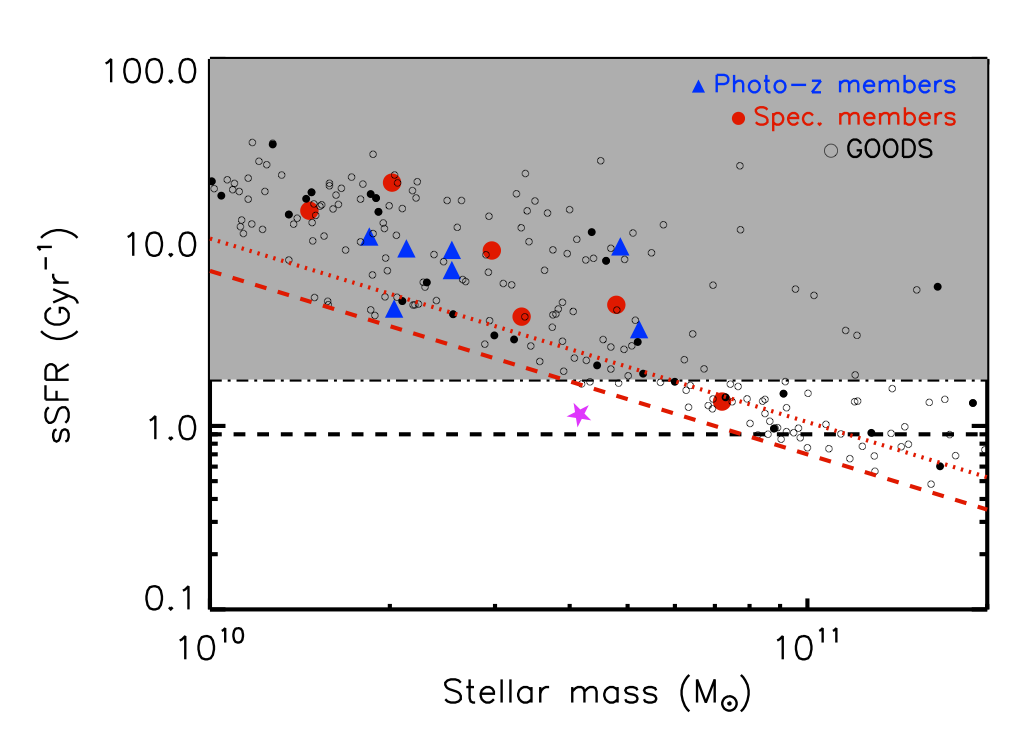} 
 \caption{sSFR as a function of galaxy stellar mass for the spectroscopic members (red), photo-z members (blue) and for GOODS (black). 
 The magenta star indicates the sSFR and average stellar mass of the stacked galaxies in Section 5.
 The dashed line corresponds to the relation found by \citet{Elbaz11} representing the locus of the main sequence galaxies at the cluster redshift. 
 The shaded area indicates the domain of starburst galaxies at $z$=1.393.
 The sSFR detection limits of our data are shown by the red lines (2-$\sigma$ in dashed line, 3-$\sigma$ in dotted line).}
\end{figure}

To understand the role of environment in star-formation at high-redshift, we compare the FIR SFRs in this 
massive, biased system, with that of non-cluster fields observed with \textit{Herschel} at the same depth and redshift.
Only the very deep \textit{Herschel} observations of the GOODS North and South fields obtained within the context of the PACS Evolutionary 
Probe \citep{Lutz} have the necessary depth to directly compare the star formation properties in low-density 
environments (i.e. fields) with respect to this cluster.
Using the most recent catalog of GOODS provided by the PEP consortium, we select a subsample of 33 field galaxies with 
spectroscopic redshifts in the range 1.35--1.45, and 255 photometric redshifts galaxies within the same redshift interval used for the cluster (i.e. 1.1--1.7)
 for which we have stellar masses, $L_{IR}$ and SFR above our 2$\sigma$ detection limit in the cluster.

We find that the field starburst galaxy population with a similar redshift of XMM2235 occupies the same 
region of our cluster in the SFR -- $M_{*}$ and sSFR -- $M_{*}$ relations, as shown in Fig. 6. 
%Perhaps more interesting here is the comparison with equally massive low-redshift clusters, that show a distinct lack of ULIRGs (see Section 4), 
%whereas in XMM2235 we have a large fraction (2/3 of the FIR sample) of ULIRGs.

\section{AGNs in the cluster}

The number and distribution of AGNs in clusters provides important clues on the growth of the supermassive black holes at the centers of clusters, 
the nature of AGN fueling, and the impact of AGNs on the intracluster medium over cosmic time. Moreover, the numerous arguments that 
support a connection between star formation and black hole accretion indicate that there should be an increased AGN population in distant clusters.
Few studies have investigated the number and evolution of AGNs in clusters in a systematic way and using a statistically significant sample 
(Galametz et al. 2009, Martini et al. 2009, and more recently Martini et al. 2013). 
Specifically, Martini et al. (2009), quantifies the evolution of the fraction of luminous X-ray AGNs in clusters
from $f_{A}$ = 0.13$^{+0.18}_{-0.087}$\% at $z$ = 0.2 to $f_{A}$ = 1.00$^{+0.29}_{-0.23}$\% at $z$ = 0.7, or more simply,
 a factor of 8 increase to $z$ = 1.
 
Using the sub-arcsecond resolution data from \textit{Chandra} we identify point source X-ray emission from AGNs associated with the cluster 
members ID 5, 237, 1227 and 2669, all of them with FIR emission. The X-ray luminosities of these AGNs span the range 
0.8--6.7 $\times$10$^{43}$ erg/s in the hard band (2.0-10 keV), that corresponds to type-2 (obscured) AGN.

We attempt to estimate the contribution of the AGNs associated with these four galaxies to the FIR luminosity by performing a 
decomposition of the SEDs with Decompir\footnote{https://sites.google.com/site/decompir/}, 
a program that separates infrared spectral energy distributions into their AGN and host-galaxy components \citep{Mullaney}.
While for ID 5, 237 and 1227, the fit appears to be entirely dominated by the host galaxy, for ID 2669 there is a hint for an AGN component 
that dominates the FIR bump. We argue that we have a tentative case for AGN contamination in this cluster member, since this 
isolated galaxy has $F_{100\mu m} > F_{160\mu m}$, which is best explained if we consider emission by an AGN.
However, given the large uncertainties of this fit caused by the poor data at 250/350/500$\mu$m, we unable to quantify the FIR emission 
due to the AGN.
Using the FIR luminosities computed with \textit{LePhare} in Section 4 we tentatively find an anti-correlation between the X-ray and far-infrared 
luminosities for these four galaxies.

\section{Morphologies}

The morphological type of a galaxy is tightly correlated with its star-formation activity \citep[e.g.][] {Wuyts, Bell}.  
Star-forming galaxies are typically associated with 
late-types (mostly spirals and irregulars) whereas early-types (ellipticals and spheroidals) - the predominant class in massive clusters -
are mostly passive, "red and dead" galaxies, at least up to $z\le$2. 
Using the high angular resolution of HST we can determine the morphological classification of these small, distant galaxies, either by
a visual comparison with reference templates (e.g. Postman et al. 2005 for high-redshift galaxies) or by fitting surface brightness models 
(e.g. Sersic, de Vaucouleurs models) to the data, and studying the distributions of the output parameters \citep[e.g.][]{Blakeslee}.  
We perform here a visual classification of the morphological type of the members with PACS detection and therefore a high star-formation rate.  
All seven photometric-redshift members have coverage with ACS, however only 2 of the 6 spectroscopic members were imaged with HST. 
We present in Fig. 7 postage stamps of these galaxies using the ACS/F850LP band. As expected, 
most if not all galaxies have a complex morphology with irregular features, typical of late-types. 
About half of the galaxies in this sample (ID 5, 157p, 225p) shows signs of on-going mergers, the preferred mechanism to explain starbursts.
We note that galaxy ID 237 (which is located very close to a bright star) shows X-ray point source emission.

\begin{table*}
\caption{Far-infrared properties of PACS selected cluster member candidates.}  
\label{table:2}      
\centering          
\small
\begin{tabular}{llllll} 
\hline\hline   
ID       & RA       & DEC                                                 &      $L_{IR}$             & SFR   & Distance   \\   
          & (J2000) & (J2000)                                         &  ($\times$ 10$^{12}L_\odot$) &   (M$_\odot$/yr)  & (arcmin)  \\
\hline                                      						
     10p  &   338.73006870  &   -25.98595720             & 3.40$\pm$0.30              & 244$\pm$50    &  6.6 \\   
     11p*  &   338.85847880  &   -25.98851850           &  2.71$\pm$0.26      &   463$\pm$44  &  2.1 \\  
     38p  &   338.87219989  &   -26.00703430            &  2.38$\pm$0.47       & 406$\pm$80    &  3.5 \\  
     44p  &   338.79876460  &   -25.93098120            &  2.27$\pm$0.33       & 387$\pm$57    &  2.9 \\ 
     45p  &   338.84994860  &   -25.93286289            & 1.80$\pm$0.34        & 309$\pm$59    &  1.9 \\ 
     49p  &   338.86812700  &   -26.00802690            & 2.21$\pm$0.37        & 376$\pm$64    &   3.4  \\ 
     53p  &   338.80206950  &   -25.97794499            & 1.41$\pm$0.32        & 242$\pm$56    &  2.3  \\  
     54p  &   338.78448880  &   -25.91753960            & 1.92$\pm$0.34        & 330$\pm$57    &  4.1  \\
     64p  &   338.84216420  &   -25.89336420            &  2.07$\pm$0.39       & 354$\pm$66    & 4.1 \\ 
     72p  &   338.90687459  &   -25.94695199            & 2.40$\pm$0.50        & 410$\pm$86    & 4.3 \\  
     78p  &   338.81335430  &   -25.93354280            & 1.48$\pm$0.43        & 252$\pm$72    &  2.2 \\  
     80p  &   338.74649740  &   -25.93000479            &  3.25$\pm$0.37       & 556$\pm$63    &  5.7  \\   
     81p  &   338.74888979  &   -25.95501159            & 1.69$\pm$0.33        & 289$\pm$57    & 5.3 \\   
     84p  &   338.81575800  &   -25.87702519            & 1.39$\pm$0.28        & 238$\pm$48    & 5.2  \\  
    103p  &   338.84948470  &   -25.91015870           & 1.50$\pm$0.29      & 256$\pm$50     & 3.1 \\  
    106p  &   338.81899199  &   -25.99635719           & 1.74$\pm$0.40      & 298$\pm$68     & 2.4 \\   
    107p  &   338.90347270  &   -25.92848229           & 2.09$\pm$0.36      & 358$\pm$61     & 4.4 \\  
    111p  &   338.85414689  &   -26.05769240           &  2.38$\pm$0.33     & 407$\pm$57     & 5.9  \\  
    116p  &   338.81277010  &   -26.00859930           & 1.45$\pm$0.35      & 248$\pm$59     & 3.2 \\ 
    128p*   &  338.86492919  &   -25.98668479          &  1.03$\pm$0.16      &  176$\pm$26  & 2.3  \\   % very close to a spec interloper and photo-z member but this is the one!
    142p  &   338.85945659  &   -25.93265829           &  1.63$\pm$0.39      & 279$\pm$67    & 2.2 \\ 
    148p  &   338.91433599  &   -25.94082639           &  1.28$\pm$0.21      & 219$\pm$37    & 4.8 \\
    153p  &   338.91098099  &   -25.99418820          &   3.2$\pm$0.30         & 142$\pm40     $   & 4.9   \\  
    160p*,**$^{+}$ =ID5 &   338.84048461  &   -25.95356750  &  1.55$\pm$0.19    &       268$\pm$34    &  0.5  \\
    162p  &   338.86938900  &   -25.92509529         & 1.35$\pm$0.28         & 232$\pm$49     & 2.9  \\  
    165p  &   338.85021430  &   -26.05912419         & 1.71$\pm$0.32         & 293$\pm$55     & 5.9 \\  
    170p  &   338.88531280  &   -25.91803669         &  0.77$\pm$0.32        & 132$\pm$56   & 3.9  \\   
    172p  &   338.77226990  &   -25.97174250         &  0.63$\pm$0.21        & 107$\pm$45   & 3.9 \\  
    181p  &   338.83815980  &   -25.89049740         &  1.58$\pm$0.33        & 270$\pm$57   & 4.2 \\  
    188p  &   338.77278940  &   -25.98506469         & 1.53$\pm$0.29         & 261$\pm$51   & 4.1  \\
    192p  &   338.90547299  &   -25.99396829         & 1.63$\pm$0.32         & 279$\pm$55   & 4.6  \\  
    208p &   338.90187320  &   -25.96328069          & 1.23$\pm$0.28         & 211$\pm$49   & 3.9 \\ 
    218p  &   338.89078660  &   -25.96478700         &  1.29$\pm$0.39        & 222$\pm$67   & 3.2  \\  
    219p  &   338.81149520  &   -25.94031409         &  0.68$\pm$0.18        & 115$\pm$31   & 2.0  \\  
    243p  &   338.85018020  &   -26.00747660         & 0.96$\pm$0.25         & 164$\pm$42   &  2.9 \\  
    246p*$^{*,+}$  &   338.85324096  &   -25.94196701      &  1.16$\pm$0.22     &   199$\pm$38   &  1.5 \\
    252p  &  338.86190795   &   -25.93078422         & 1.51$\pm$0.44          & 259$\pm$75          &   2.4  \\
    300p*$^{+}$  &  338.84832763   &  -25.93917846         &  1.16$\pm$0.33      &   198$\pm$56 &   1.5 \\  % photo-z ID 1195
    301p*  &  338.85226440   &  -25.95386314         &    0.52$\pm$0.21      &    89$\pm$35     &  1.0  \\    % photo-z ID 1356
    304p$^{+}$  &  338.85748291   &  -25.94927978  &   0.95$\pm$0.28       & 162$\pm$47   &  1.4 \\  % the source centered on PACS does not have a spec or photoz 
    327p  &   338.75933609  &   -25.98157600         &    1.47$\pm$0.27       & 251$\pm$47   &  4.8 \\ 
\hline  
\end{tabular}
\flushleft  \hspace{3.cm}   * photo-z member, ** spectroscopic cluster member, $^{+}$ narrow-band emitter
\end{table*}

\section{Candidate cluster members selected in PACS data}

In addition to the study of the sources for which we have redshift information (either from spectroscopy or photometric redshifts), 
we exploit the full field-of-view of the PACS images to perform a more general selection of FIR sources that are likely to be highly star forming candidates. 
This selection follows from the assumption that the peak of the SED is at $\sim$ 240$\mu$m for a starburst at the cluster redshift 
(see e.g. Krikpatrick 2012, Casey 2012), which allows us to define the simple color criterion F250 $>$ F160 $>$ F100 $\mu$m. We chose not to 
add the condition F350 $<$ F250 in our selection, due to the poor resolution of SPIRE at long wavelengths.

A total of 60 sources (of the 158 3-$\sigma$ sources in the matched 100-160$\mu$m catalog) meet our selection criteria.
Of these, eleven PACS candidates are found to be spectroscopic interlopers, three more are photo-$z$ interlopers, and after a visual screening of the 
individual galaxies in the PACS and the Ks-band data we discard 5 additional sources which are considered spurious or for which we have a low S/N. 
 Of the remaining 41 sources listed in Table 4,  one is the spectroscopic member ID 5, and  
 six have a photometric redshift within a range of 0.15 from the cluster redshift.
  
We note that not all cluster members listed in Table 2 are included in this sample, since for most of those cluster galaxies 
we do not have 3-$\sigma$ fluxes in both PACS bands. Nonetheless, 
we confirm that generally their SEDs satisfy our flux selection criterion, with the exception of ID 2669, which has 
F$_{160}$=4.9$\pm$1.24  mJy and F$_{100}$= 5.82$\pm$0.45 mJy. This is likely explained by the presence of an AGN component 
as shown in Section 7, which boosts the 100 $\mu$m flux. 

To estimate the contamination in this sample we use the photometric redshift catalog, which is less prone to selection biases.
We find 10 photo-$z$ galaxies in the initial catalog of 60 PACS selected sources, of these 6 are photo-$z$ member candidates, and 4 are
photo-$z$ interlopers, ie., their redshift is outside the range [1.1--1.7]. Two of the 4 photo-$z$ interlopers are also spectroscopic interlopers. 
Even though this is a rough estimate, we consider a contamination of 40\% in our sample of purely infrared selected cluster candidates.
This number may be higher though, because the photometric redshifts have an expected fraction of spurious sources of 50\%.

All candidates are located outside the cluster core ($r>$250 kpc) and $\sim$85\% are beyond $R_{500}$ in the range of 1--3 Mpc from the BCG. 
Since we do not have photometric redshifts for a large part of this sample as most of our optical data is limited to the inner few arcminutes, we assign 
to all candidates the cluster redshift and perform a FIR SED fitting analysis to obtain the total LIR and SFR. The results of this
analysis are listed in Table 4. 
The best approach to confirm this sample is to have near-infrared spectroscopy follow-up.

The PACS candidates and H$\alpha$ emitter catalogs overlap within a very small FOV (see Fig. 8)  that includes only 9 PACS candidates. 
Of these, 4 galaxies are present in both catalogs. The remaining 5 PACS candidates that are not seen in H$\alpha$ may be either interlopers or
 strongly obscured.% ($A_{H\alpha} \sim$ 3 mag). 
The sensitivity of the \textit{Herschel} maps allows us to detect only the high star-forming galaxies, 
so we miss the low SFR galaxies that constitute the bulk of the H$\alpha$ detections. 
  ~Several studies have reported high ($A_{H\alpha} \sim$ 3 mag) extinction levels when 
comparing mid- or far-infrared sources with H$\alpha$ emitters with similar stellar mass in high redshift clusters \citep[e.g.][] {Koyama}.
In the following section we estimate the extinction in H$\alpha$.

\begin{figure}
\includegraphics[width=8.cm,angle=0]{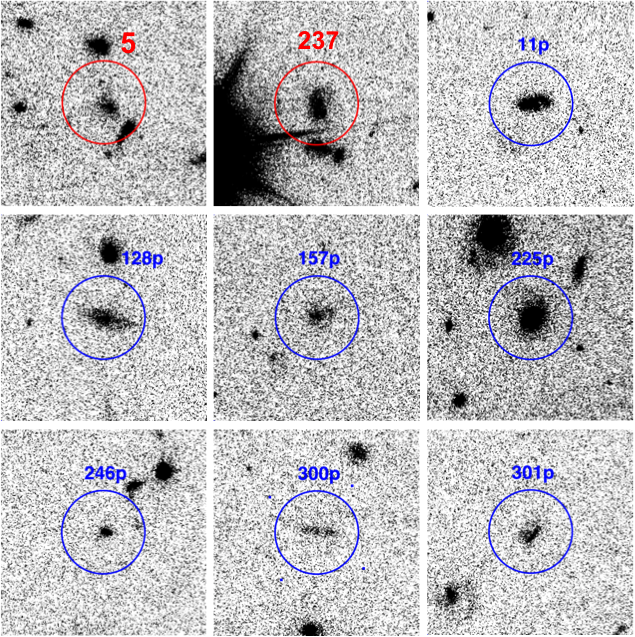} 
 \caption{HST/ACS-F850 10$\arcsec \times$10$\arcsec$ stamps of 2 spectroscopic cluster members (ID 5, 237) and 7 photo-$z$ members 
  of XMM2235, used to assess the morphologies of these galaxies with emission in PACS. The circles have a radius of 2\arcsec.}
\end{figure}

\subsection{SFR(H$\alpha$) vs. SFR(FIR)}

We compare the star-formation rates obtained via the H$\alpha$ luminosity as presented by \citet{Grutzbauch}, 
with our SFRs measured by integrating the far-infrared luminosities, $L_{IR}$, and applying the \citet{Kennicutt} law. To perform this
comparison we use a set of 6 galaxies with both SFR(H$\alpha$) and SFR(FIR) measurements, for which we have either a spectroscopic 
or a photometric redshift. This is a small sample in part because the overlapping field-of-view is small ($\sim$ 5$\arcmin \times$3.5$\arcmin$),
and because the depth of our \textit{Herschel} images corresponds into a minimum star-formation rate of the order of 70 (2-$\sigma$) -- 105(3-$\sigma$) M$_\odot$/yr, not allowing us to detect low star-forming galaxies.

We find that the quoted SFR(H$\alpha$) is systematically lower than the corresponding SFR(FIR), SFR(H$\alpha$)$\sim$0.1-0.2 SFR(IR).
The amount of extinction in H$\alpha$ needed to reconcile these two independent star-formation diagnostics is 3 magnitudes, a 
reasonable value for the range of SFR spanned by our galaxies \citep[e.g.][] {Koyama}.  
Also recent work by Ibar et al. (in preparation) found a similar trend when comparing the SFR(H$\alpha$) with star-formation rates obtained from 
\textit{Spitzer}-MIPS or \textit{Herschel} using a sample of field galaxies at $z$=1.47.

\section{SINFONI IFU data of the BCG: upper limit on H$\alpha$ emission}

XMM2235 is currently the most distant cool-core cluster on the basis of its X-ray properties (i.e., surface 
brightness concentration $c_{SB}$, short cooling time and a hint of a temperature drop towards the core) 
that were studied in detail using deep \textit{Chandra} data in \citet{Rosati}. 
Besides these well-known X-ray properties, the cool-core phenomenon sometimes also manifests in the central 
galaxy through star-formation caused by the inflow of cold gas in the cluster center. 
Indeed, \citet{Cavagnolo} demonstrated that local clusters with low central entropy
 (which is closely related to the central cooling time) derived from their X-ray surface brightness properties 
 have a high probability to have a BCG with optical line emission. 
 However, the effect of star formation induced by cooling in otherwise red, passive elliptical 
galaxies is very challenging to measure, with reported fractions of BCGs with [OII] emission of the order of 35\% in local cool-cores \citep{Donahue}. 
At high-redshifts measuring this effect is even more challenging due to the faintness of distant galaxies. Moreover, we may 
observe and measure X-ray quantities that show that a cluster is a cool-core, however, the triggering of SF in the BCG may be delayed and thus absent.

The measurement of star-formation in the central galaxies of cool-core clusters using \textit{Herschel} is limited to low redshift clusters 
(e.g. Mittal et al. 2012) due to the high star-formation thresholds imposed by the \textit{Herschel} data. Still, Rawle et al (2012) recently 
conducted a study in BCGs up to $z$=1 and found that star-forming BCGs require a small ($<$1 mag) extinction correction to 
reconcile SFR(H$\alpha$) with SFR(FIR) which should be linked to a different source of fuel (cold gas) other than the normal stellar mass loss.

The BCG of XMM2235 is a very large and massive elliptical galaxy ($\sim$10$^{12}M_\odot$) 
perfectly centered with respect to the peak of the X-ray emission.
We were awarded with a 7h Integral Field Unit (IFU) spectroscopy of the BCG using the near-infrared spectrograph SINFONI 
at ESO/VLT \citep{Eisenhauer, Bonnet}, to investigate the presence of H$\alpha$ emission in the central galaxy as a tracer of 
star formation (PI Santos, 383.A-0825A). The observations were taken in August 2009, in non-AO mode with an on--source exposure
of 5 hours in the 8$\arcsec \times$8$\arcsec$ field.
We performed a standard data reduction and obtain the final spectrum by median stacking of the individual spectra. 

We do not detect any convincing H$\alpha$ emission, hence we report a
3-$\sigma$ upper limit on the H$\alpha$ flux in the XMMJ2235-25 BCG of
9$\times10^{-17}$erg/s/cm$^{2}$. This translates to an upper limit on the star-formation rate 
of 7.6 M$_\odot$/yr for Kennicutt (1998) with a Salpeter initial mass function, assuming a line width of 150km/s 
and no reddening correction.  This upper limit is a factor 10 lower than the one we can place with the current \textit{Herschel} data.

\begin{figure*}
\includegraphics[height=6.5cm,angle=0]{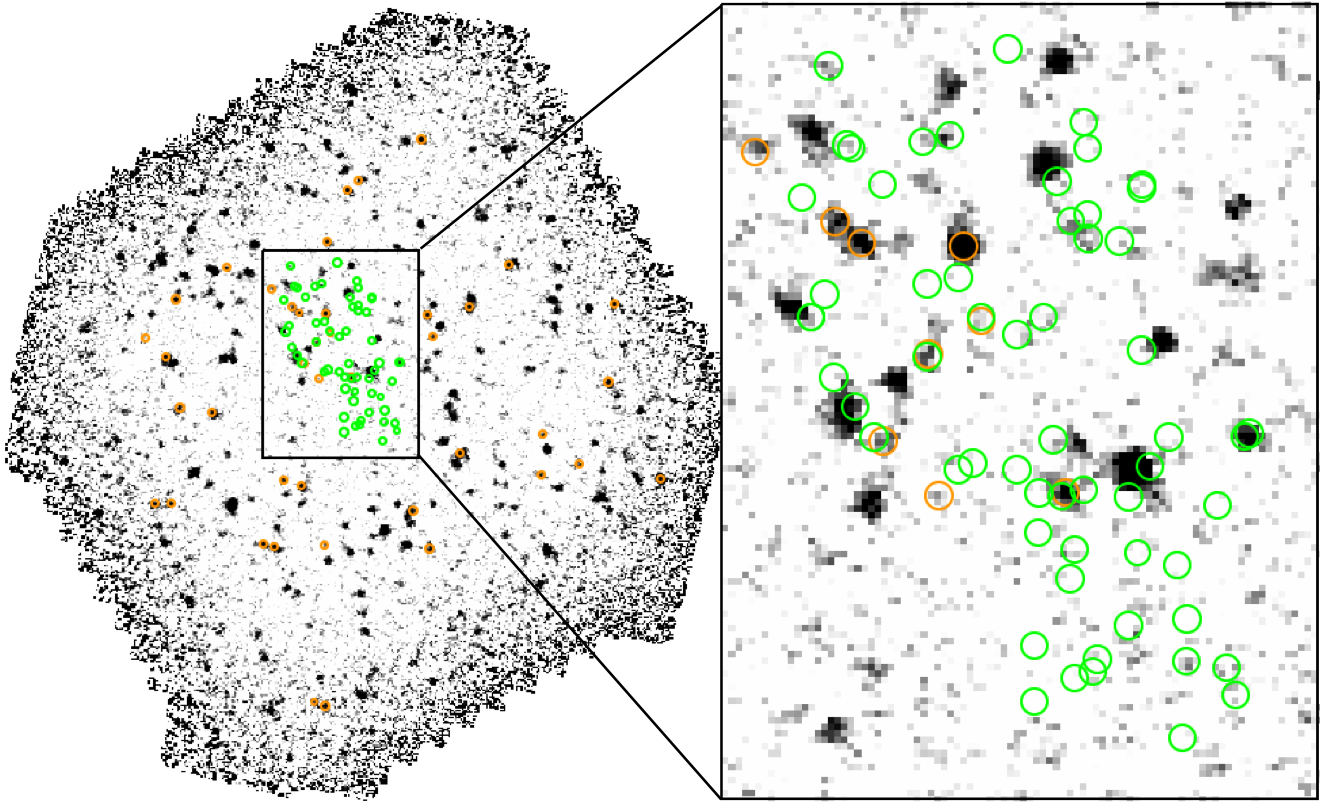} 
 \caption{PACS selected cluster member candidates (orange) and H$\alpha$ emitters (green). The black square indicates the overlapping FOV 
 with 3$\arcmin \times$4 $\arcmin$. }
\end{figure*}

\section{Conclusions}

In this paper we have characterized the far-infrared properties of the galaxies (confirmed and photo-$z$ candidates) 
in XMM2235, a massive cluster at $z$=1.4, using deep \textit{Herschel} imaging data from 100-500 $\mu$m.
The extensive ancillary data of this cluster is crucial to ensure a proper identification of the cluster galaxies in the \textit{Herschel} maps, 
and to further characterize the properties of the highly star-forming members.
We summarize our main results here:

\begin{enumerate}
  \item we derived star-formation rates for the 13 spectroscopic and photo-$z$ members in the range 89--463  M$_\odot$/yr, 
  all of them lie outside the core at $r>$ 250 kpc from the  center and extend well beyond $R_{500}$;
     \item we obtained a significant stacking signal of 9 spectroscopic members that were not individually detected by PACS, corresponding to a 
  SFR = 48$\pm$16 M$_\odot$/yr;
  \item we find that 4 of the 6 spectroscopic members detected by PACS are associated with AGN X-ray emission, although the data did not 
  allow for a decomposition of the FIR emission into an AGN and host components;
  \item we performed a visual morphological classification of a subsample of nine members with HST/ACS data which show that these high 
  star-forming galaxies are all late-types;
  \item  we find that SFR (H$\alpha$) is a factor 5--10 lower than SFR (IR), which indicates that a dust extinction of about 3 magnitudes 
 in H$\alpha$;
\item the sSFR--M$_{*}$ relation shows that all cluster members with FIR emission are starburst galaxies that follow the same trend as field galaxies 
at the same redshift;
  \item we selected a sample of 41 FIR cluster member candidates using a color criterion based on the shape of the SED. To confirm these are 
  highly star-forming members we require follow-up observations in the near-infrared.
\end{enumerate}

We also presented deep IFU SINFONI data of the BCG, with the aim to detect H$\alpha$ emission since this is 
  a cool-core cluster. We place a strict upper limit of 7.6 M$_\odot$/yr to the H$\alpha$ derived star-formation, which is 
  much below the upper limit we can place using the FIR data. The lack of H$\alpha$ emission in the massive, large, central galaxy implies that the 
  star-formation induced by the cold gas in the core has not started yet (since this is a very distant cluster) or already happened on a short timescale.

Even though this is currently the deepest \textit{Herschel} data of a distant galaxy cluster, the FIR study of the individual cluster members is 
 still quite challenging because the poor resolution of the data makes it hard to disentangle contamination from undesired objects.
Furthermore, only highly star-forming galaxies can be detected with this data, i.e. our 2$\sigma$ (3$\sigma$) limit on the SFR is 
70(105) M$_\odot$/yr, allowing us to detect mostly ULIRGS. Stacking cluster members proved to be an important resource 
to go below the nominal SFR limit and reach the main-sequence level. Nevertheless, follow-up observations in the near-IR 
(ideally multi-object spectroscopy) targeting the H$\alpha$ emission line covering the large FOV of PACS are needed to confirm the
 cases where there is significant contamination and to confirm the large sample of PACS selected cluster galaxy candidates. 

At $z$=1.4 we do not witness a reversal of the SFR--density relation in this massive cluster. As shown in Fig. 4, all cluster members with 
FIR emission and PACS selected candidates are located at a radius greater than 250 kpc from the cluster center.
Six of the thirteen individual detections are beyond a radius of 1 Mpc, approximately the cluster virial radius, therefore most of the 
measured FIR star-formation in this cluster occurs in potentially infalling galaxies at the edge of the cluster X-ray emission.
We note, however, that current studies providing evidence for an increase of SF in the centers of distant clusters 
used MIPS 24$\mu$m data and thus had a lower limit to the star-formation rate. 

The study presented here on the individual system XMM2235 is part of a larger project including more than a dozen galaxy clusters reaching the 
proto-cluster regime, that will allow us to characterize star-formation activity at different cosmic times and stages of evolution.

\section*{Acknowledgments}

JSS thanks D. Elbaz and R. Gobat for useful discussions
and R. Gr\"utzbauch for sending us the yet unpublished catalog of revised SFR of the H$\alpha$ emitters.
R.D. acknowledges the support provided by the BASAL Center for Astrophysics and Associated Technologies and by FONDECYT N. 1100540.
C.L. is the recipient of an Australian Research Council Future Fellowship (program number FT0992259). 

PACS has been developed by a consortium of institutes
led by MPE (Germany) and including UVIE (Austria); KUL, CSL,
IMEC (Belgium); CEA, OAMP (France); MPIA (Germany); IFSI, OAP/AOT,
OAA/CAISMI, LENS, SISSA (Italy); IAC (Spain). This development has been
supported by the funding agencies BMVIT (Austria), ESA-PRODEX (Belgium),
CEA/CNES (France), DLR (Germany), ASI (Italy), and CICYT/MCYT (Spain).

SPIRE has been developed by a consortium of institutes
led by Cardiff Univ. (UK) and including Univ. Lethbridge
(Canada); NAOC (China); CEA, LAM (France); IFSI,
Univ. Padua (Italy); IAC (Spain); Stockholm Observatory
(Sweden); Imperial College London, RAL, UCL-MSSL,
UK ATC, Univ. Sussex (UK); Caltech, JPL, NHSC, Univ.
Colorado (USA). This development has been supported by
national funding agencies: CSA (Canada); NAOC (China);
CEA, CNES, CNRS (France); ASI (Italy); MCINN (Spain);
SNSB (Sweden); STFC and UKSA (UK); and NASA (USA).

Based on observations made with the European Southern Observatory (ESO) 
telescopes at Paranal Observatories under programme ID 383.A-0825(A), 274.A-5024(B), 
077.A-0177(A, B), 074.A-0023(A), 077.A-0110(A, B).

%\appendix

%\section[]{Large gaps in L\lowercase{y}${\balpha}$ forests\\* due to fluctuations in line distribution}
%
%\newpage
%
%\subsection{Subsection title}

%We plot in Fig.~\ref{appenfig} $P(>x_{\rmn{gap}})$ for several $N$
%values. We see that, for $N=100$ and $x_{\rmn{gap}}=0.06$,
%$P(>0.06)\approx20$ per cent.  This means that the probability of

\bsp

\label{lastpage}

\end{document}